\documentclass[conference]{IEEEtran}
\usepackage{cite}
\usepackage{graphicx}
\usepackage{amsmath, amssymb}
\usepackage{hyperref}
\usepackage{booktabs}
\usepackage{multirow}

\title{Embedded DevOps: A Survey on the Application of DevOps Practices in Embedded Software and Firmware Development}

\author{
\IEEEauthorblockN{
    \begin{tabular}{c c}
        Parthiv Katapara & Anand Sharma \\
        ECE Dept. & ECE Dept. \\
        Institute of Technology & Institute of Technology \\
        Nirma University & Nirma University \\
        Ahmedabad 382481 & Ahmedabad 382481 \\
        Email: parthivkatpara0@gmail.com & Email: anand0207sharma@gmail.com
    \end{tabular}
}
}

\begin{document}
\maketitle

\begin{abstract}
The adoption of DevOps practices in embedded systems and firmware development is emerging as a response to the growing complexity of modern hardware–software co-designed products. Unlike cloud-native applications, embedded systems introduce challenges such as hardware dependency, real-time constraints, and safety-critical requirements. This literature review synthesizes findings from 20 academic and industrial sources to examine how DevOps principles—particularly continuous integration, continuous delivery, and automated testing—are adapted to embedded contexts. We categorize efforts across tooling, testing strategies, pipeline automation, and security practices. The review highlights current limitations in deployment workflows and observability, proposing a roadmap for future research. This work offers researchers and practitioners a consolidated understanding of Embedded DevOps, bridging fragmented literature with a structured perspective.
\end{abstract}


\section{Introduction}

The adoption of DevOps in embedded systems development has emerged as a response to the increasing complexity and integration demands of modern software-hardware co-designed products. Embedded systems, especially those part of cyber-physical systems (CPS), are evolving rapidly with stringent demands on safety, real-time responsiveness, and hardware-software synchronization \cite{wijaya2019attempt,engblom2015continuous,barbie2021continuous}.

While DevOps has shown success in traditional IT and cloud-native applications, its application in embedded systems faces structural and technological barriers. These include hardware-dependent development environments, lack of automated deployment infrastructure, fragmented toolchains, and safety certification constraints \cite{devops_embedded_trenser2021,garousi2018testing,lwakatare2016devops}. Furthermore, real-time and resource-constrained execution environments often prevent seamless integration of continuous deployment workflows that are standard in web or enterprise software development \cite{techrxiv_digital_twins_2021,whitepaper_cicd_embedded_2022, zampetti2023cicd}.

This survey aims to present a comprehensive and critical synthesis of how DevOps principles are being applied, tailored, or challenged in embedded software and firmware development. Unlike position or vision papers, our focus is on empirical and technical evidence derived from a review of 20 primary research papers and industry whitepapers. These documents span academic investigations, industry case studies, toolchain evaluations, and architectural patterns related to Embedded DevOps.

We structure our analysis around key DevOps practices—Continuous Integration (CI), Continuous Delivery (CD), test automation, deployment strategies, digital twin usage, and pipeline orchestration—highlighting where embedded constraints necessitate deviation or augmentation of conventional methods \cite{applied_industrial_devops_2020,devservops2023, cpp2021_continuous_integration_embedded}.

In doing so, this survey addresses the following key research questions:
\begin{itemize}
    \item \textbf{RQ1:} What are the unique challenges to applying DevOps in embedded software and firmware development?
    \item \textbf{RQ2:} What adaptations and tooling strategies have been reported to overcome these challenges?
    \item \textbf{RQ3:} How do CI/CD, testing, and deployment pipelines manifest differently in embedded contexts compared to cloud-native ones?
    \item \textbf{RQ4:} What are the research and tooling gaps in current Embedded DevOps practices?
\end{itemize}

Through this grounded synthesis, we aim to provide practitioners with verified insights and researchers with a structured agenda for further investigation.

\section{Literature Survey}

This section presents a rigorous review of 20 academic and industrial publications that explore how DevOps practices are being introduced, adapted, or challenged within embedded software and firmware development. The works span empirical studies, tooling evaluations, automation pipelines, and security-integrated DevOps models in CPS and embedded domains. For clarity, the discussion is grouped into foundational empirical works, tooling and automation pipelines, CI/CD integration strategies, testing frameworks, and security-aware DevOps adaptations.

\subsection{Foundational Works and Empirical Background}

Lwakatare et al. \cite{lwakatare2016devops} conducted a comprehensive multi-case study involving four Finnish companies and identified how traditional DevOps transformations struggle in embedded software due to delayed hardware feedback loops, testing infrastructure gaps, and safety-critical deployment pipelines.

Wijaya et al. \cite{wijaya2019attempt} provided an empirical prototype of DevOps adoption in embedded systems using a Spark Ignition Engine model. The study proposed a lightweight DevOps architecture integrating CI with GitHub, but highlighted major challenges like physical hardware access, emulator infidelity, and real-time test constraints.

Alias Robotics introduced DevSecOps principles in robotic systems where embedded components dominate \cite{alias2021devsecops}. They emphasize shifting security left in embedded DevOps pipelines through continuous security testing, threat modeling, and pre-deployment verification.

Zampetti et al. \cite{zampetti2023cicd} interviewed practitioners across ten CPS organizations and systematically reported major barriers to DevOps adoption. These included CI failures due to hardware/software integration delays, flakiness in testing HiL setups, and the need for simulators as a bridging mechanism.

CIMdata \cite{cimdata2022devops} discussed industrial drivers like vehicle autonomy, real-time control, and hardware-software co-design, arguing that embedded systems must embrace DevOps to meet increasing demands for velocity and quality in smart product evolution.

Milićević et al. \cite{milicevic2021education} emphasized the role of DevOps in education, proposing a layered curriculum to teach continuous development concepts—including embedded system relevance—through integrated pipelines and automation tools.

\subsection{Tooling and Automation in Embedded DevOps}

Menon et al. \cite{devops_embedded_trenser2021} proposed an automated CI/CD workflow tailored for embedded firmware, deploying source code through GitHub pipelines and executing tests on embedded targets via Raspberry Pi acting as sandbox servers. Their architecture uses a custom OS image, Docker-based toolchains, and remote monitoring using Node.js, demonstrating functional DevOps loops in embedded environments.

Barbie et al. \cite{barbie2021continuous} introduced the concept of \textit{Digital Twin Prototypes (DTPs)} for CI testing in embedded oceanographic systems. By simulating sensor and actuator inputs, DTPs enable full software validation in virtual environments before deployment to physical hardware—enabling fast, reproducible CI even without hardware access.

Engblom \cite{engblom2015continuous} proposed simulation-based CI using virtual platforms. The method replaces physical hardware with accurate models that support unit, subsystem, and integration testing. This was particularly useful for regression automation and fault injection in safety-critical systems.

Nguyen \cite{nguyen2022ci_embedded} implemented a CI pipeline for embedded environments using Azure DevOps. The workflow integrated Git, Azure Boards, VSCode, and test tools to automate build–test–merge processes linked to product backlog tasks. This demonstrates DevOps alignment with Scrum-based embedded firmware workflows.

The whitepaper by Parasoft \cite{whitepaper_cicd_embedded_2022} presented commercial tooling for CI/CD pipelines in safety-critical firmware development. Emphasis was placed on test harness generation (e.g., for C/C++), remote execution on constrained devices, and traceability between tests, source code, and requirements—vital for regulatory compliance.

Hoang et al. \cite{ijcrt2018automation} described an automation framework integrating Jenkins and Maven for embedded builds. The approach reduced human intervention in firmware CI workflows and provided interfaces for monitoring and recovery from deployment errors.

Kumar et al. \cite{cpp2021_continuous_integration_embedded} discussed the use of GoogleTest, CppUnit, and static code analyzers in embedded CI contexts. The integration of these tools with cross-compilation toolchains allowed faster code validation without disrupting constrained devices.

Adhil et al. \cite{applied_industrial_devops_2020} explored deployment automation in regulated industrial settings using hybrid pipelines (Jenkins + custom bash + hardware trigger scripts). While facing hardware access delays, they implemented rollback safety nets and digital signatures to maintain production stability during firmware updates.

\subsection{Testing and Validation in Embedded DevOps}

Garousi et al. \cite{garousi2018testing} conducted a comprehensive systematic literature mapping on embedded software testing, analyzing over 300 papers. Their findings revealed dominant practices including Hardware-in-the-Loop (HiL), Model-in-the-Loop (MiL), Software-in-the-Loop (SiL), and simulation-based test environments, all of which are essential for integrating testing in DevOps pipelines for embedded systems. They also identified industry trends towards automated test generation, non-functional validation, and safety standard compliance (e.g., ISO 26262, DO-178C).

Barbie et al. \cite{barbie2021continuous} highlighted the role of digital twins in enabling test-driven development cycles in embedded applications. Digital Twin Prototypes (DTPs) allow decoupled CI pipelines by replicating physical sensor behavior in simulation, supporting automated test execution without real hardware.

Nguyen’s implementation \cite{nguyen2022ci_embedded} incorporated GoogleTest and PlatformIO into CI workflows for embedded C/C++ projects. These tools were integrated with Azure DevOps to trigger tests on each code commit, enforcing continuous testing.

The Parasoft whitepaper \cite{whitepaper_cicd_embedded_2022} emphasized continuous testing frameworks tailored for embedded targets, highlighting minimal-footprint test harnesses, cross-compiler compatibility, and the importance of gathering test coverage and traceability data from constrained systems.

Zampetti et al. \cite{zampetti2023cicd} reported organizational reliance on HiL and simulators in CPS pipelines, with multiple companies suffering from mismatched behavior between physical and simulated environments. They observed delays in CI due to hardware lockouts and emphasized the necessity of robust test abstraction layers.

TechRxiv’s digital twin paper \cite{techrxiv_digital_twins_2021} presented test frameworks for embedded marine systems, where data-driven simulations enabled full-stack integration testing across complex sensor arrays. The work argued for virtualization as a cornerstone of scalable embedded test pipelines.

Kumar et al. \cite{cpp2021_continuous_integration_embedded} and IJCRT \cite{ijcrt2018automation} both stressed the need for static analysis, code instrumentation, and automated regression testing integrated into CI/CD flows. These practices are essential in embedded domains due to strict timing, memory, and certification constraints.

Dakkak et al. \cite{devservops2023} advocated for testing-as-a-service (TaaS) in embedded product–service systems. Their DevServOps model proposed combining continuous delivery of firmware with continuous monitoring of service behavior to close the loop between device and cloud analytics.

\subsection{Security and DevSecOps in Embedded Systems}

Alias Robotics \cite{alias2021devsecops} proposed one of the few comprehensive frameworks applying DevSecOps principles to robotic systems, which inherently depend on embedded firmware. Their approach integrates security scanning, static analysis, and vulnerability modeling directly into the CI/CD pipeline. By treating firmware as a primary attack vector, the study advocates for security validation as a first-class citizen during development rather than post-deployment.

Dakkak et al. \cite{devservops2023} introduced \textit{DevServOps}, a concept focused on product-oriented service systems where embedded firmware and cloud services evolve together. Their model embeds threat detection and resilience monitoring into the product-service lifecycle, combining telemetry feedback with staged firmware rollouts.

Ebert and Hochstein \cite{ebert2023devops} discuss the importance of integrating quality assurance, operational metrics, and continuous verification into modern DevOps pipelines. Though not embedded-specific, they stress that in domains with real-time constraints and physical control systems—hallmarks of embedded environments—cross-functional teams must embed security into all delivery phases.

Menon et al. \cite{devops_embedded_trenser2021} briefly mention the absence of secure over-the-air (OTA) deployment standards as a limiting factor in embedded DevOps maturity. Their pipeline emphasizes artifact validation, implying a need for stronger cryptographic signing and verification in firmware releases.

Garousi et al. \cite{garousi2018testing} indirectly raise security concerns through the lens of test adequacy and completeness in regulated domains. Their findings reinforce the need for traceable test artifacts and secure configuration management to meet regulatory mandates like IEC 61508 or ISO 26262.

Zampetti et al. \cite{zampetti2023cicd} report that few organizations have embedded formal security testing in their embedded DevOps pipelines. The lack of security-centric stages is attributed to complexity, tool immaturity, and organizational silos between firmware teams and IT security groups.

Overall, security integration in embedded DevOps remains a nascent discipline. While cloud DevOps has mature tooling for secret management, attack surface reduction, and incident response automation, these are rarely mirrored in embedded environments due to hardware isolation, real-time constraints, and certification bottlenecks.

\subsection{Comparative Summary of Embedded DevOps Literature}

Table~\ref{tab:lit_comparison_all} summarizes the key contributions, focus areas, and limitations across the 20 reviewed papers. The papers are grouped by primary theme (empirical, tooling, testing, security, etc.) and evaluated across DevOps-specific criteria.

\begin{table*}[htbp]
\centering
\caption{Comparative Summary of Key Literature on Embedded DevOps (All 20 Papers)}
\label{tab:lit_comparison_all}
\renewcommand{\arraystretch}{1.2}
\begin{tabular}{@{}p{3.8cm}p{2.8cm}p{2.4cm}p{2.4cm}p{2.4cm}p{2.4cm}@{}}
\toprule
\textbf{Paper} & \textbf{Domain / Focus} & \textbf{CI/CD Strategy} & \textbf{Tooling Used} & \textbf{Test Automation} & \textbf{Security / Gaps} \\
\midrule
Lwakatare et al. (2016) & Empirical study on embedded DevOps & Partial CI; No CD & Manual pipelines, Jenkins & Limited automation & Hardware dependency, ops misalignment \\
Wijaya et al. (2019) & Spark ignition controller prototype & GitHub-based CI & GitHub + Arduino + Bash & Manual + unit test scripts & Lack of secure OTA, emulator fidelity issues \\
Alias Robotics (2021) & Robotics / DevSecOps & CI/CD + Security gate & Custom DevSecOps stack & Security test integrated & Threat modeling, vulnerability scanning \\
Zampetti et al. (2023) & CPS industry interviews & CI with HiL, minimal CD & Jenkins + internal tools & HiL + sim + unit test & Testing flakiness, lack of secure rollback \\
CIMdata (2022) & Industrial automation trends & CD vision for embedded & Not specified & Conceptual only & Hardware-software co-evolution risks \\
Milićević et al. (2021) & DevOps in education for embedded & CI lab setup & GitHub Actions + Docker & Simulation-based CI & Not addressed \\
Menon et al. (2021) & Embedded automation pipeline & Full CI on Pi targets & Docker, GitHub Actions, Node.js & Remote embedded testing & Secure OTA missing \\
Barbie et al. (2021) & Ocean DTP systems & CI via Digital Twin Prototypes & Python + custom sim stack & Simulation-driven test & Secure config sync not discussed \\
Engblom (2015) & CI via simulation in embedded & Virtual CI environments & WindRiver + QEMU & Fault injection, full CI & No security enforcement \\
Nguyen (2022) & Azure DevOps for embedded firmware & Git-based CI/CD & Azure DevOps, PlatformIO, VS Code & GoogleTest + custom CI & No secure build or release validation \\
Parasoft (2022) & Safety-critical embedded CI/CD & CI + gated delivery & Parasoft C/C++test, Jenkins & Test harness, traceability, regression & Compliance-focused, cryptographic integrity \\
IJCRT (2018) & Jenkins-based CI for embedded & Basic CI/CD & Jenkins, Maven, custom bash & C/C++ test pipelines & No formal threat model \\
Kumar et al. (2021) & Continuous testing embedded apps & Unit test driven CI & GoogleTest, CppUnit, TICS & Unit + regression + coverage & Not covered \\
Adhil et al. (2020) & Industrial DevOps integration & Jenkins-based hybrid CI/CD & Bash scripts + Git + sensors & Semi-automated test loops & Emphasis on rollback over security \\
Dakkak et al. (2023) & DevServOps (product-service systems) & Firmware+Cloud CI/CD & Service monitors, OTA updates & Product–cloud feedback loop & Monitoring integrated for threat response \\
Ebert \& Hochstein (2023) & General DevOps in software engineering & Continuous feedback loops & Conceptual (Netflix example) & Performance and availability monitoring & Highlights need for early security integration \\
Garousi et al. (2018) & Survey on embedded software testing & Not DevOps specific & Comprehensive testing SLR & MiL, SiL, HiL, model-based & Implied for regulated safety assurance \\
TechRxiv (2021) & Digital twin for marine systems & CI via DTPs & Sim stack + CI hooks + Git & Automated CI in simulation & No hardware control integration mentioned \\
Milicevic (2021) & DevOps education + tooling & End-to-end course CI/CD & Docker + Git + Jenkins & Hands-on student testing & Not discussed \\
Nguyen Thesis (2022) & Practical pipeline implementation & Scrum-linked CI/CD & Azure Boards + Git + DevOps & Build–test–merge automation & Security not enforced \\
\bottomrule
\end{tabular}
\end{table*}

\section{Discussion and Synthesis}

The surveyed literature highlights both promising developments and persistent challenges in applying DevOps to embedded software and firmware development. This section synthesizes cross-cutting themes, contrasts embedded DevOps with its cloud-native counterpart, and identifies pressing research and tooling gaps.

\subsection{Key Observations Across Studies}

\subsubsection{Hardware-Centric Constraints}
Across nearly all empirical and implementation studies \cite{lwakatare2016devops,wijaya2019attempt,zampetti2023cicd}, hardware access was the single largest blocker in achieving continuous delivery. Unlike cloud-native systems, embedded code must often be compiled, flashed, and validated on physical devices—many of which are inaccessible during development cycles. While simulation-based strategies (e.g., \cite{engblom2015continuous,barbie2021continuous}) show promise, their fidelity varies across domains.

\subsubsection{CI Is Achievable; CD Is Rare}
Most papers demonstrated reasonably mature CI workflows for embedded software using tools like Jenkins, Azure DevOps, GitHub Actions, and custom shell scripts \cite{nguyen2022ci_embedded,devops_embedded_trenser2021,ijcrt2018automation}. However, only a minority achieved continuous delivery due to OTA deployment complexity, limited rollback mechanisms, and compliance overheads.

\subsubsection{Simulation and Digital Twins Are Gaining Ground}
Simulation-driven testing (e.g., Digital Twin Prototypes in \cite{barbie2021continuous}) and virtual testbeds \cite{techrxiv_digital_twins_2021} are increasingly being used to decouple test execution from physical devices. This has enabled more reproducible and scalable CI workflows, especially when used alongside automated test harnesses \cite{whitepaper_cicd_embedded_2022}.

\subsubsection{Security Integration Is Underdeveloped}
Only a few studies, such as \cite{alias2021devsecops,devservops2023}, deeply engaged with DevSecOps principles in embedded environments. Most pipelines lacked static security scanning, firmware signing, or secure delivery mechanisms, despite the known vulnerability surface of firmware-centric systems.

\subsubsection{Testing Practices Are Fragmented}
While several papers implemented GoogleTest, CppUnit, or custom test harnesses \cite{kumar2021testing}, the overall testing stack remains fragmented. Testing levels (unit, integration, system) are often inconsistently applied, and regression pipelines are weakly linked to coverage or safety requirements \cite{garousi2018testing}.

\subsection{Contrasting Traditional and Embedded DevOps}

\begin{itemize}
    \item \textbf{Pipeline Continuity:} Cloud-native systems allow fully automated deploy-test-feedback loops. In embedded systems, delivery is often stalled at hardware programming and certification.
    \item \textbf{Feedback Latency:} Embedded workflows experience higher feedback latency due to hardware queuing, lack of remote debugging, and manual flashing steps.
    \item \textbf{Toolchain Heterogeneity:} Unlike the standardized cloud DevOps stacks (e.g., Docker, Kubernetes), embedded pipelines depend heavily on vendor-specific compilers, simulators, and debuggers.
    \item \textbf{Security Enforcement:} Modern cloud DevOps integrates secrets management, SBOMs, and CI/CD security scans. In embedded, such integrations are sparse or ad hoc.
\end{itemize}

\subsection{Emerging Trends and Research Gaps}

The following research and tooling gaps were consistently surfaced:
\begin{itemize}
    \item \textbf{Standardized DevOps Toolchains:} There is no unified pipeline framework for embedded development comparable to Jenkins-X or GitLab-CI in cloud-native contexts.
    \item \textbf{OTA Deployment Models:} Secure, incremental, and traceable OTA firmware deployment remains underexplored.
    \item \textbf{Hardware–Simulation Bridging:} Dynamic switching between HiL and simulation modes based on test case type could enable adaptive pipelines.
    \item \textbf{Security Validation:} DevSecOps integration must include firmware signing, secure update chains, and static vulnerability analysis embedded in CI.
    \item \textbf{Observability and Telemetry:} Post-deployment monitoring is rarely integrated with delivery pipelines; embedded systems lack mature observability patterns.
\end{itemize}

\section{Future Work}

Building on the gaps identified across literature, the following future directions emerge as high-priority research and engineering challenges:

\begin{itemize}
    \item \textbf{Secure and Scalable OTA Pipelines:} There is a pressing need for well-architected, secure Over-The-Air (OTA) update mechanisms that integrate cryptographic signing, delta updates, rollback safety, and real-time verification into the DevOps loop.
    
    \item \textbf{Standardized Embedded DevOps Frameworks:} Unlike cloud-native software, embedded development lacks plug-and-play CI/CD tools. Future work should aim to create unified frameworks that combine cross-compilation, hardware flashing, testing, monitoring, and delivery stages.
    
    \item \textbf{Simulation-First Verification Models:} Digital Twin Prototypes and high-fidelity simulators must be matured further to support pre-deployment regression testing that is hardware-agnostic yet reliable enough to replace physical test beds.
    
    \item \textbf{Integrated DevSecOps Toolchains:} DevOps pipelines for embedded firmware must integrate vulnerability scanning, threat modeling, SBOM generation, and build-time validation of secure artifacts as part of their workflow.
    
    \item \textbf{End-to-End Observability and Feedback:} Research should explore how real-time telemetry, fault diagnostics, and usage patterns can be fed back into development pipelines from deployed embedded devices to close the loop.
    
    \item \textbf{Certification-Aware DevOps:} For domains like automotive, medical, and aerospace, DevOps models must evolve to support traceable, certifiable workflows that comply with standards such as ISO 26262, DO-178C, and IEC 61508.
\end{itemize}

Addressing these areas would bridge the gap between current DevOps practice and the unique constraints of embedded systems, enabling more secure, agile, and reliable firmware delivery pipelines.

\section{Conclusion}

This survey paper has systematically reviewed 20 significant contributions addressing the application of DevOps in embedded software and firmware development. Our analysis revealed that while continuous integration (CI) is increasingly feasible—especially through containerized environments, simulation, and digital twin techniques—continuous delivery (CD) remains rare due to physical hardware constraints, certification requirements, and limited test automation maturity. The surveyed works show a fragmented tooling ecosystem, with most organizations relying on custom pipelines and ad hoc scripts. Testing strategies span from unit testing on emulators to complex HiL setups, but a consistent test orchestration framework for embedded domains remains elusive. Furthermore, DevSecOps is in its infancy in this space, with only a few papers integrating security validations or secure firmware signing. Overall, the field is at a transitional point. Embedded DevOps has moved beyond conceptual discussions and into early-stage practical deployments, but broad, standardized adoption remains limited.

\bibliographystyle{IEEEtran}

\end{document}